# Mediated behavioural change in human-machine networks: exploring network characteristics, trust and motivation

Paul Walland and J. Brian Pickering

IT Innovation Centre
Gamma House, Enterprise Road
Southampton, SO16 7NS, UK
`{pww,jbp}@it-innovation.soton.ac.uk`

**Abstract.** Human-machine networks pervade much of contemporary life. Network change is the product of structural modifications along with differences in participant behavior. If we assume that behavioural change in a human-machine network is the result of changing the attitudes of participants in the network, then the question arises whether network structure can affect participant attitude. Taking citizen participation as an example, engagement with relevant stakeholders reveals trust and motivation to be the major objectives for the network. Using a typology to describe network state based on multiple characteristic or dimensions, we can predict possible behavioural outcomes in the network. However, this has to be mediated via attitude change. Motivation for the citizen participation network can only increase in line with enhanced trust. The focus for changing network dynamics, therefore, shifts to the dimensional changes needed to encourage increased trust. It turns out that the coordinated manipulation of multiple dimensions is needed to bring about the desired shift in attitude.

## 1    Introduction

Human-machine networks[1] (HMNs) pervade all aspects of contemporary life from social and family interactions to retail, online learning and eDemocracy, or to managing investments, tax and even visa applications. These HMNs display varying characteristics, offering many different ways to interact and achieve whatever the goals of participants might be [1]. But they can and do change as users alter their behaviours [2]; and attempts to account for such changes often fail to appreciate user expectation and creativity when they engage with the network, especially when explanations of dynamic change are based on a reductionist notion of networks as theoretical models only [3]. A more pragmatic approach might be to weigh the expectations and actions of network users, participants in the activity that the network was set up for or how it has evolved, as expressed in terms of their own perceptions and interactions with the network against the characteristics of the network design. In this paper, we will present an approach to eliciting user goals in connection with a citizen participation network and contrast this with indicators derived from a profile of the network from the HUMANE project (https://humane2020.eu/). Taking citizen participation as an example HMN, we will show that the tension between the top-down controls imposed by network design and the bottom-up expectations of user goals facilitates the identification of issues and challenges which need to be addressed if the network is to be successful in meeting the needs of participants and maintaining their trust and motivation.

---

[1] We use the terms human-machine network (HMN) and network interchangeably.

## 2  Citizen participation networks

Given the reach of the Internet [4], it is no surprise that human-machine networks are also present within democratic processes [5, 6]. However, even though social networks are well-known globally, it is not clear whether activity in social networks necessarily encourages participation in democratic practices [7]. Indeed, on- and offline democratic processes do differ [4]: people are used to social networks and online debate, but this does not translate directly into participatory behavior [8]. eDemocracy and eParticipation will therefore tend to complement rather than replace traditional processes [5]. HMNs extend the reach of debate without improving its quality, and may even reduce participation down to selected groups only [9]. That being so, the assumption that eParticipation is about political decision-making may be false. Instead, online discussion may simply lead to a more refined understanding of a single issue [10]. So an HMN that encourages discussion and the frank exchange of views between citizens may well be more appropriate than one hosting interactions with politicians and policy makers. Citizens are therefore given a chance to discuss with one another about the local issues that affect them directly rather than having to consider long-term solutions to national issues. This is not difficult to understand: online discussion with peers is really an inherently social activity [10] and will be influenced, therefore, by social forces [11, 12]. In consequence, HMNs supporting citizen participation have to integrate social, political and technical factors if they are to succeed [5, 13]. Citizens may, therefore, want such a network to provide an opportunity to discuss issues that affect them with their peers. However, it is equally possible that policy makers and politicians want access to the electorate to gauge their response to planned legislation. For either set of stakeholder, the network platform may or may not be able to provide the technical capabilities which both needs. There is therefore an inherent challenge in striking a balance between stakeholder expectations on the one hand, and socio-technical issues such as acceptability, system adoption and willingness, on the other [13]. That being the case, how might we characterise a human-machine network and identify what needs to be done to support these complex socio-technical requirements?

## 3  Profiling human-machine networks

**Fig. 1** summarises a set of eight characteristics or dimensions grouped into four abstract layers. Each of the dimensions has been given a relative value in this figure just to illustrate a broad-stroke representation of the network: there is more human agency than machine agency, for example, since machines simply aggregate input whilst human agents can interact with each other in different groups, moderate each other's views and inputs, and formulate opinions and strategies together. Similarly, geographic and network size are comparable since an eParticipation network is at least theoretically open to everyone within a given geographic location. That location may expand up to national level, for instance, like the UK government survey site https://petition.parliament.uk/. Independently, the dimensions allow individual aspects of the network to be examined; further, taken within an individual abstract layer, the interplay between the pair of dimensions would allow additional investigation of network dynamics [14, 15]. How this dynamism affects the network offers the opportunity to control and manipulate the network. In so doing, if there are issues in a human-machine network, it would be possible that these limited dimensions provide a mechanism to modify the network and encourage behavioural change [16].

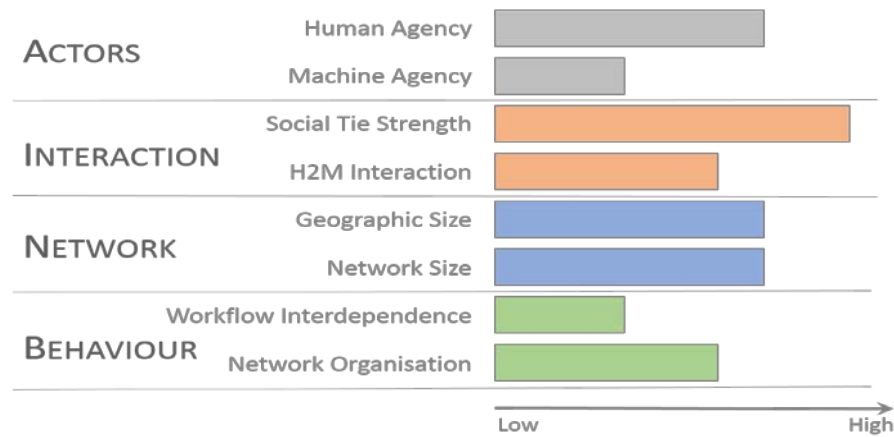

**Fig. 1.** Dimensions of a citizen participation human-machine network

As shown in **Fig. 1**, the individual dimensions across the four abstract layers of HMNs may be understood as follows: **actors**, comprising *Human* and *Machine agency* to reflect the levels of capability of the different actors; their **interactions,** or *Social Tie Strength* representing the relationship between human agents and *H2M Interaction* between human and machine agents; the **network**, which may expand over a large *Geographic* area such as *facebook*, or a more modest regional group like *Mumsnet*, and which may include many actors (a substantial *Network size*) or just a few; and lastly **behaviour**, referring to *Workflow interdependence* (how independent actors in the network are to do what they want), and *Network organisation* (whether imposed from above or top-down; or emergent from the actions of members themselves, i.e., bottom-up) [17]. Within any given abstract grouping, such as *actors*, individual dimensions may be more or less independent of one another. For instance, *Human agency* may well increase along with *Machine agency* as the capabilities of the machine components allow the human agents to do more and more things [14]. Manipulating these dimensions as we suggest will, of course, affect this, either opening up additional opportunities or restricting others. The co-ordinated increase in Human and Machine agency just mentioned, for instance, is likely to increase participants' perception of self-efficacy: they can do more and possibly achieve more. Similarly, reducing Network organisation together with Workflow interdependence would increase autonomy for HMN participants. However, there may be other implications. Increasing autonomy in this way may lead to confusion about what can and cannot be done in the network, and therefore, reduce self-efficacy compromising motivation to engage with the network. So the HUMANE typology based on a limited set of dimensions can result in manipulations to the structure of the network. If such changes affect attitudes, this structural change may in turn help actors in the network achieve their ultimate goals. At the same time, though, it is important to consider how those dimensions might affect generic, cross-cutting concerns such as motivation, trust and participation. To identify these intangible issues, or *meta-dimensions*, we may need to consider more than the structure of the network and its characteristics. Engaging directly with participants to establish what they need and expect from the HMNs they participate in will help clarify how network dimensions may be used as controls in modifying network outcomes.

## 4    Identifying stakeholder perspectives

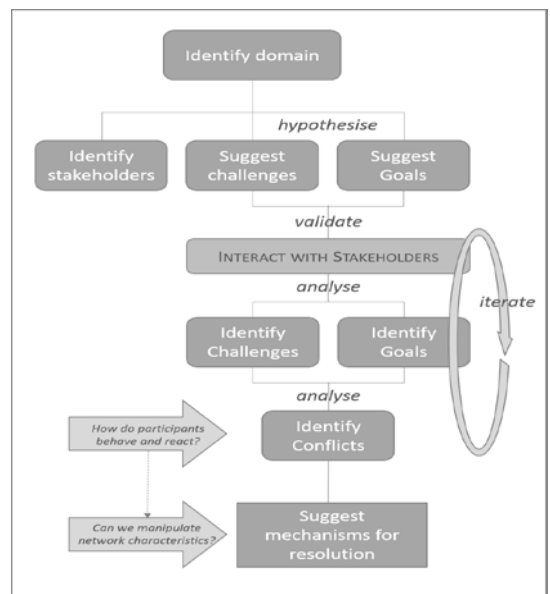

**Fig. 2.** Exploring participant perceptions

So, returning to the example of eParticipation, stakeholders for the HMN may need to include policy makers, politicians, citizens and lobby groups to try and establish what their priorities are. It is possible to derive or surmise their respective goals from observation, discussion and questionnaire. It turns out they may well have conflicting objectives: the goals of the policy makers and politicians (which may or may not be the same) seem to be to increase acceptance of the legislation they propose, while the citizens themselves and the lobby groups representing them may want to define the type and scope of legislation for politicians to implement. To understand those expectations as they develop and are shaped by ongoing change in the social and political context, we need a suitable iterative process. This will also allow potential participants to explore the domain and the network itself. **Fig. 2** encapsulates just such a process.

Having identified the stakeholders relevant to the domain, we can begin by suggesting what goals those stakeholders aspire to. Citizens, for instance, want their voices to be heard, and to be acted on. Policy makers want to understand what is important to the citizens and how they will respond to proposed legislation. Lobbyists want to ensure that their point of view reaches the policy makers, and that their objectives (and those of their members) are met. At the same time, we might recognise that there are blockers or challenges to achieving those goals. Citizens may feel constrained to express what they truly feel, or lack trust that what they say will be taken into account. Policy makers may need to decide between opposing views as well as different interests in formulating policy. And lobbyists need to demonstrate that they have faithfully represented those they serve *and* that there has been a positive effect. (For a discussion of these issues, see [18].)

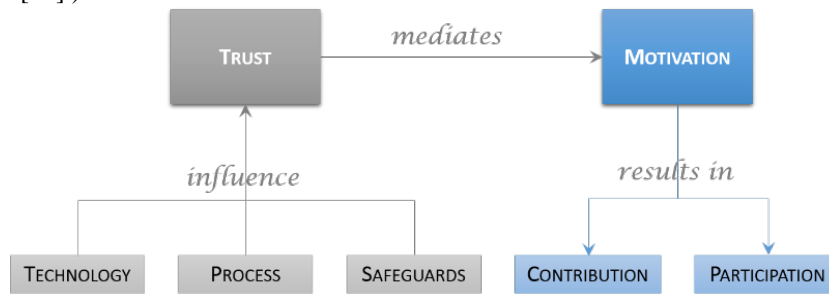

**Fig. 3.** A simple model of meta-dimensions in citizen participation HMNs

Validating these challenges and goals with the stakeholders allows us to home in on how participants react. In monitoring their behaviours, we can identify what the relevant meta-dimensions for this network are. In a recent study, a small group of self-selecting actors involved in citizen participation took part in an online survey [19]. First they ranked stakeholder roles in order of importance within the network, resulting in the following rank order: Citizen groups (most important), Non-Government Organisations (NGOs), Government, Policy makers, and least important, IT Professionals and designers. It is perhaps surprising to find Government and Policy makers ranked below Citizens themselves and NGOs. However, this reflects some of the observations from the literature whereby citizen engagement is more about debate and social interaction than necessarily establishing contact with Policy makers [9, 10, 13].

Turning to goals and barriers, respondents went on to identify and without ranking them in order the main goals of citizen participation networks to be: (i) *Managing trust*; (ii) *Generating a culture of engagement*; (iii) *Encouraging open and transparent debate*; (iv) *Motivating participation from all parties*; and (v) *Accountability*. These relate particularly to **trust in the network** and **motivation**. Trust may be affected by a number of different aspects of the network such as the technology involved and how competent individual users feel with the technology. Respondents had voiced some concern leading to the identification of one of the challenges to network success would be understanding the real role of technology. Alternatively, looking at accountability and a call for open and transparent debate, trust is affected by perceptions of the process itself: does participation really make a difference, for example? Here again, one obstacle to achieving the overall network goals was the desire to see outcomes being publicised and made available to all actors in the network. Finally, trust will be affected by whatever safeguards are in place to ensure accountability and protect open discussion.

Trust therefore is influenced by the factors summarised in **Fig. 3**. In itself, trust is not available for manipulation: you cannot make someone trust a process or another person. However, it is possible to change the factors technology, process and safeguards. In this way, changes in trust as a consequence of manipulation of these factors will have an effect on motivation. The figure shows this as the mediating influence of trust, and the underlying drivers of technology, process and safeguards, in encouraging motivation. Stakeholders identified the need to generate a culture supportive of debate and to encourage participation and contributions from citizens. Motivation is precisely about driving active participation so that citizens and citizen groups provide input and contribute to the success of the network. So the way to facilitate these beneficial effects of motivation, we need to consider how we might affect the influencing factors which lead to changes in trust.

**Fig. 3** therefore presents a simple model of two meta-dimensions: trust and motivation. At the top level, trust and motivation are key components within the network reflecting what influences behaviour and how that behaviour contributes to the success of the network. The traditional definition of trust as a willingness to expose oneself to vulnerability [20–22] would predict motivation to be present so long as there is some understanding of possible risk and what regulation might be in place to mitigate against that risk. Motivation itself nevertheless might be made up of the willingness to participate, echoing the social aspects of engaging in public debate [10, 12]. But at the same time, there is a desire to contribute, show competence and affiliation, and demonstrate commitment to a cause [23, 24]. This simple model is based on the overall goals and related challenges for citizen participation networks. The question arises as to whether the dimensions we discussed above could inform what needs to be done in the network to address the challenges and move the HMN towards the overall goals identified by participants. In other words, how might the dimensions of the network (**Fig. 1**) engender trust in participants which would in turn lead to motivation? Clearly, the dimensions would only have an indirect influence on trust itself via the constructs technology, process and safeguards. That being the case, individual network dimensions may affect other constructs related to network behaviours which will need further investigation.

To begin with, though, we need to decide the consequences of changing individual network dimensions in terms of the meta-dimensions proposed.

## 5    Manipulating networks through controlling network dimensions

Given the simple model of citizen participation networks, encouraging trust in the HMN so that this will influence and improve motivation is not straightforward. Although the network dimensions summarised in **Fig. 1** offer one way to manipulate outcomes in the network, as stated above, they do not directly affect trust. Instead, they must be applied to components in the network (technology), the outcomes of the network (process) and external factors influencing the network (safeguards). Only then will a propensity for participants to trust become possible. Starting with the first set of dimensions, *actors* covers both Human and Machine agency. It is clear that increasing machine agency in support of human agency provides a good opportunity to facilitate self-efficacy and a sense of competence increasing motivation and thereby a willingness for participation. If we then turn to *interactions* in the network, increased Social Tie Strength between human actors, mediated by machines, will facilitate communication and promote publication of outcomes and transparency as identified by citizen participant stakeholders in our survey. By contrast, raising H2M interaction strength will improve self-efficacy providing human actors know-how to use the technology which provides the applications and services required. As self-efficacy increases so trust in the overall process might be expected to improve. With regard to *network,* the question of Network size and scale demands particular consideration. The geographical scope of a regional or national eParticipation network might be constrained to include only those geographically affected by outcomes of such participation (e.g. just within a national jurisdiction), whilst within that geographical scope, as many human nodes as possible should be enabled to participate. If the digital divide is an issue, then the network must minimise it: the technology must be implemented with ease of use and user experience the most important design principle. Finally, *behaviour* refers to what can and cannot be done as a result of the network. So Workflow interdependence should be carefully considered, since it hinges on whether social norms, distributed via peer engagement, are enabled to a sufficient extent to improve or lower perceived risk. This can be regarded as a helpful safeguard, which would lead to increased trust. This is probably where the interplay between different actors (e.g. policy makers and citizens) is most significant, as well as where trust in the process is greatest. As a result of this interplay, the system of government is created. Finally Network organisation, in the sense of top-down *versus* bottom-up organisation, should be considered. In this case, a bottom-up evolution of a network doesn't necessarily lead to participation in governance, as both *facebook* and *Twitter* have shown. Rather than peer-to-peer private sharing, they now tend to be used for commentary and broadcasting concern. Similarly, top-down creation of participation networks by government have also failed to achieve success, since there is insufficient trust in the system, and poor communication of the objectives of the network. Yet a purely bottom-up approach would lack the cohesion required to support constructive debate, and exposes the network to side-tracking by extremism. There is therefore a need to consider what the identification of an 'intermediate' organisation might be, and how a network that is neither bottom-up nor top-down might be created.

So the abstract layers and network dimensions which the HUMANE project proposes provide a way to alter network characteristics. Looking at the potential effects if changing individual dimensions, means that we might predict how to change network dimensions in order to bring about behavioural change. For, as the characteristics themselves are modified, so the behaviours associated with the network would be constrained or freed up to engage in different types of activity. In the simple model of citizen participation networks in **Fig. 3**, it is how the network dimensions within different abstract layers affect individual network components like the technology, the processes the network is designed to support, and the necessary constraints and safeguards put in place to secure the network and its processes. If the model is correct, then the resulting increase in trust would in turn alter motivation. As suggested, this indirect outcome – by manipulating network dimensions, which change trust levels, which in turn mediates motivation, leading to greater participation and so forth – reveals a process for behavioural change in networks which deserves more attention. What we need to do now, though, is go one step further in trying to identify the effects of multiple dimension changes at the same time.

## 6 Trust and agency in human-machine networks

In a recent study on how trust, human agency and machine agency might affect behaviour, Pickering et al. [15] modified previous work by Thatcher, McKnight and their colleagues on trust in technology [25, 26]. **Fig. 4** shows the original research model Pickering et al. proposed (with individual constructs in black), and is centred on the relationship between trust and behaviour. Briefly, *regulation* controls what can and cannot be done in a network, that is, it has a direct effect on human and machine agency; similarly, regulation will provide input to the estimation of any *perceived risk*. Together, they act as safeguards associated with the network. Already, it is clear that influencing individual network components in the hope of promoting trust may not be so straightforward. Interestingly, as regulation is typically external (linked with legislation and similar controls), so perceived risk tends to be internal (the result of some form of cognitive algebra). It is this internal safeguard – an individual's notion of what is and isn't risky – which affects trust and network behaviours directly, not the external safeguard, regulation. *Machine agency* affects *human agency*, i.e., it allows human agents to achieve more in the network, and thereby increases *self-efficacy*, the perception of what an individual is able to do. *Self-efficacy* will in turn affect network behaviours, along with trust in the network and *social norms*. This model tests the effects we suggest above that can be produced on the network by manipulation of the network dimensions as outlined. But what it is already showing is that increasing trust is not simply about the aggregated effect of changes in technology, process and safeguards.

Taking this one stage further (the shaded constructs in **Fig. 4** with labels in italics), linking safeguards and agency with trust, self-efficacy and thereby network behaviours means that the original model may usefully be extended. The abstract layers of *interaction* and *behaviour* (see **Fig. 1**) affect different parts of the trust-behaviour model. Beginning with the behaviour layer, *Network organisation* can be expected to influence social norms: a bottom-up configuration would leave social cohesion outside the scope of the network, and therefore dependent solely on the individuals themselves. A top-down structure would tend to impose some degree of uniformity on how human agents can interact and communicate with one another: the network would therefore constrain the possible effects of social norms. By contrast, *Workflow interdependence* will affect both *Human* and *Machine agency*: the more structured the workflow, the greater the machine agency, but at the same time, human agency is constrained. With less structure, however, the opposite is not necessarily true. First, human agency will, of course, tend to increase: there is more flexibility in what the human agents can do. This may not lead to reduced machine agency. Instead, the machine agency may need to increase to be able to support the greater scope available to the human actors within the network.

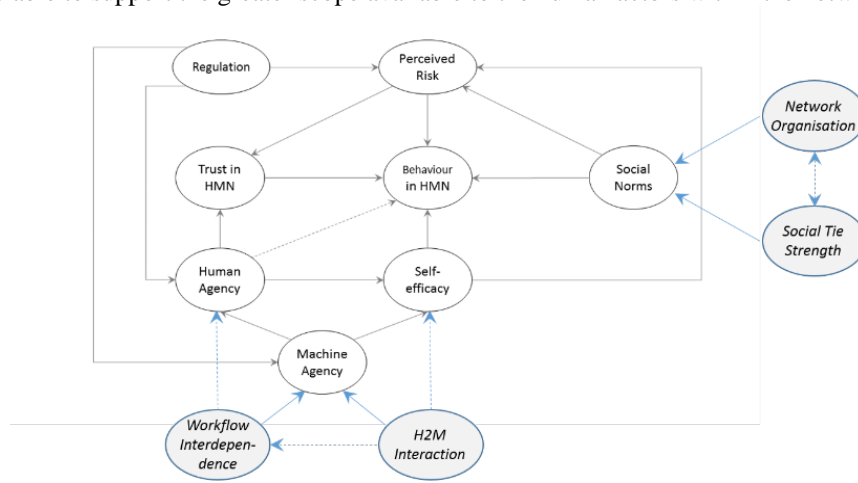

**Fig. 4.** Trust in Human-machine Networks

Turning to the interaction layer, *Social tie strength* like *Network organisation* affects social norms: as individuals identify with other players in the network, so social norms will be determined by the group identity. With weaker ties, social norms will be less influential since social identity among network users is less likely. At the same time, *H2M interaction* will influence *Self-efficacy*: as human network actors become more comfortable with what the machine components do and how they might exploit those technical capabilities, so their perception of what they can achieve will increase. Once more, though, this may require increased *Machine agency* to support the greater complexity and variety of activities which the human actors may choose to pursue. It should be noted, though, that there is a link between *Workflow interdependence* and *H2M interaction* on the one hand and *Social tie strength* and *Network organisation* on the other. First, for H2M interaction to have a significant, positive influence on Self-efficacy, Workflow interdependence would need to be less structured to open up the potential for greater Human agency and thereby increased Self-efficacy. Similarly, increased Social tie strength would suggest less top-down Network organisation.

What we see here is how important it is to manipulate multiple network dimensions at the same time to bring about structural change in the network. Coordinated network dimension change may therefore be essential if the goal is to increase trust for its mediating effect on behaviour in the HMN, which in the case of citizen participation is motivation. As motivation increases, so other behaviours will follow such as increased participation and contributions (see **Fig. 3**). At the same time, though, the relationship between network dimensions themselves begins to change. It is not enough to change a given dimension in isolation but also others may need to be controlled in tandem so that the structural characteristics of the network will produce the desired effects. Behavioural change in human-machine networks is therefore not a simple issue of providing incentives, but a more complex interaction of network factors and characteristics as they influence participant attitudes (in this case trust) which in turn affects their behaviour within the network or other non-deterministic behaviour [27].

## 7 Concluding remarks

In the discussion above, relevant stakeholders involved with citizen participation identified trust and motivation as the main issues for eParticipation. Trust is a socio-cognitive construct inaccessible to direct influence. Lack of trust was perceived as a blocker to the future development and growth of a citizen participation HMN, so that increasing it would benefit the network. In addition, trust is seen as a goal in itself: for the network to succeed and continue to be successful there has to be trust between participants and the process associated with the HMN. It can be assumed that appropriate safeguards are in place to encourage open debate and interaction. Further, trust is known to influence technology adoption [25]. As such, it may be assumed that increasing trust would be of benefit to network acceptance. Stakeholders did not identify technology acceptance directly as important for network success. Instead, they suggested aspects of internal as well as external motivation to be key to the success for citizen participation. We therefore investigated how trust might influence motivation. In so doing, the perceived blockers to achieve stated stakeholder goals for the network would be alleviated.

To investigate whether this is indeed the case, we have used a network typology. This comprises a set of dimensions which relate to different characteristics of the network. However, changing these dimensions also provides an opportunity to modify how the network operates. We examined different dimensions and what effect changing them might produce. Relating them specifically to the antecedents of trust and to possible behaviours in a network suggests mechanisms to address perceived issues in a given HMN. What we have shown based on our citizen participation example is how such an approach might apply in practice. The focus for changing network dynamics, therefore, shifts to the dimensional changes needed to encourage increased trust. It turns out that the coordinated manipulation of multiple dimensions is needed to bring about the desired shift in attitude. Finally, the network dimensions can also be shown to influence a more generic trust network.

Extending this discussion, a typology to describe network state based on multiple network characteristics or dimensions makes it possible to predict behavioural outcomes in the network. However, this work must now be developed further. Directly modifying network dimensions represents a top-down intervention, an intentional manipulation which ironically may well undermine any beneficial effects. If network participants are aware that there is an attempt to influence them, say by financial incentive, then they may well withdraw co-operation: offering rewards, for example, for altruism fails to recognise that altruism is not about external reward [28, 29]. Similarly, though, acknowledging network participant objectives without modifying the HMN to be able to support those objectives – the bottom-up imposition of participant aspirations – may simply lead to frustration and reduce motivation. How to resolve the tension between the top-down and bottom-up aspects of HMN change is the challenge we are now looking to address. What we need to do now is investigate how the top-down restructuring of the network and the bottom-up aspirations of its human actors can be reconciled.